\documentclass{osa-article}

\journal{oe}

\usepackage{booktabs}
\usepackage{cleveref}
\usepackage{ifthen}
\usepackage{subfig}
\usepackage{soul}

\newcommand{\diopter}{D}

\newcommand{\object}{O}
\newcommand{\fixationObject}{\object_f}
\newcommand{\fixationDepth}{\distance_f}
\newcommand{\testObjectA}{\object_{t1}}
\newcommand{\testObjectDepthA}{\distance_{t1}}
\newcommand{\testObjectB}{\object_{t2}}
\newcommand{\testObjectDepthB}{\distance_{t2}}

\newcommand{\nothing}[1]{}

\newcommand{\shortcite}{\cite}

\newcommand{\emacsquote}[1]{{``#1''}}

\newcommand{\new}[1]{{#1}}
\newcommand{\delete}[1]{}

\crefname{figure}{Fig.}{Figs.}
\Crefname{figure}{Fig.}{Figs.}

\begin{document}

\title{Eccentricity effects on blur and depth perception\delete{ in Light Field Displays}} 

\author{Qi Sun,\authormark{1} Fu-Chung Huang,\authormark{2} Li-Yi Wei,\authormark{1} David Luebke,\authormark{2} Arie Kaufman,\authormark{3} and Joohwan Kim\authormark{2,*}}

\address{\authormark{1}Adobe Research, Adobe Inc., San Jose, CA 95110, USA\\
\authormark{2}NVIDIA Research, NVIDIA Corporation, Santa Clara, CA 95051, USA\\
\authormark{3}Department of Computer Science, Stony Brook University, Stony Brook, NY 11794, USA}

\email{\authormark{*}sckim@nvidia.com} 


\begin{abstract}
Foveation and focus cue are the two most discussed topics on vision in designing near eye displays. Foveation reduces rendering load by omitting spatial details in the content that the peripheral vision cannot appreciate; Providing richer focal cue can resolve vergence-accommodation conflict thereby lessening visual discomfort in using near eye displays. We performed two psychophysical experiments to investigate the relationship between foveation and focus cue. The first study measured blur discrimination sensitivity as a function of visual eccentricity, where we found discrimination thresholds significantly lower than previously reported. The second study measured depth discrimination threshold where we found a clear dependency on visual eccentricity. We discuss the results from the two studies and suggest further investigation.
\end{abstract}

\section{Introduction}
\label{sec:pilot_study}
The density of retinal ganglion cells is highest in the fovea and monotonically decreases towards the periphery, causing high spatial-resolution only in the center of the visual field \cite{Kwon:2019:LBR}.
This visual effect has been harnessed by gaze-contingent (foveated) computer rendering \cite{Guenter:2012:FG,Patney:2016:TFR} \new{and imaging \cite{Tan:2018:FIN}} techniques.
These foveation methods are designed for displays with fixed focal distances, such as desktop monitors\new{, or stereo displays which offer binocular depth cues}\delete{ and VR HMDs}.
\new{In the natural world, however, objects appear at different depths. Even without stereo, the monocular eye can discriminate depths by re-focusing (accommodation). During this process, we perceive depth by comparing blur differences \cite{Cholewiak:2018:CCB}.}
Lack of accommodation depth cue may cause vergence-accommodation conflicts for binocular viewing, such as VR/AR head-mounted displays.
The conflict can be reduced by a variety of optical, display, and computation systems \new{\cite{Tan:2018:PMM,Zhan:2018:PMD}}, such as single movable focal plane, multiple fixed focal planes, or light fields \cite{Sun:2017:PFL}.
\new{These systems may require mechanical movements (e.g., movable focal plane(s) \cite{Padmanaban:2017:OVR}), optical components (e.g., multiple focal planes \cite{Zhan:2019:WMM}), or extra computation (e.g., light field rendering that causes latency)}\delete{, resulting in latency that can aggravate simulator sickness}. 

Since defocus blur provided by these displays is a key cue for monocular depth perception \cite{Read:2012:VPU} and influence vergence-accommodation conflict in binocular vision, in this paper, we ask:
{\em when viewed monocularly, do optical blur and depth perception show similar eccentricity effect as spatial resolution across the visual field?}
To help answer this question, we conducted psychophysical experiments under two display systems.
First, we measured the sensitivity of human viewers to the change in optical blur. We found that the discrimination threshold varied significantly across individuals. The measured thresholds were significantly lower than previously reported.
Second, we used a light field display (that can generate accurate and natural retinal blur without extra lens) \cite{Sun:2017:PFL} to measure depth discrimination threshold, which increased as a function of visual eccentricity. Interestingly, depth discrimination threshold changed more systematically with eccentricity than blur discrimination threshold did.
We discuss the discrepancy between blur and depth discrimination and suggest future research.

\nothing{
Our goal in rendering foveated light field is to sample the 4D information as succinctly as possible: use the fewest number of rays to represent objects at different depths and eccentricities without the user noticing the differences.
\nothing{Reduction in spatial bandwidth leads to narrower angular bandwidth as suggested by \nothing{the prior frequency analysis }\Cref{eq:retinal_angular_bound}.}%
Additional reduction in angular bandwidth may be possible if the visual system is not sensitive to changes in focal cue in the periphery.
An extreme case is to replace the peripheral 4D light field rendering with a 2D billboard, but the sensitivity and detection thresholds need to be found.
The ultimate judge for this approach is\nothing{, obviously,} the human visual system.
Specifically, we ask: are there conditions in which we can omit angular sampling without the viewer noticing it?
We designed and conducted two main sets of experiments by varying eccentricities.
In \Cref{sec:pilot_study:blur} we first start the investigation of blur detection and discrimination thresholds with optical stimuli.
However, the thresholds measured by optical blur can be more conservative than overall depth perception when the human visual system reconstructs images from a discrete 4D light field display, which has more limited bandwidth (see \Cref{sec:display_bound_analysis}).
}%

\section{Blur discrimination and visual eccentricity}
\label{sec:pilot_study:blur}

\begin{figure}[!th]
    \centering
    \includegraphics[height=7 cm]{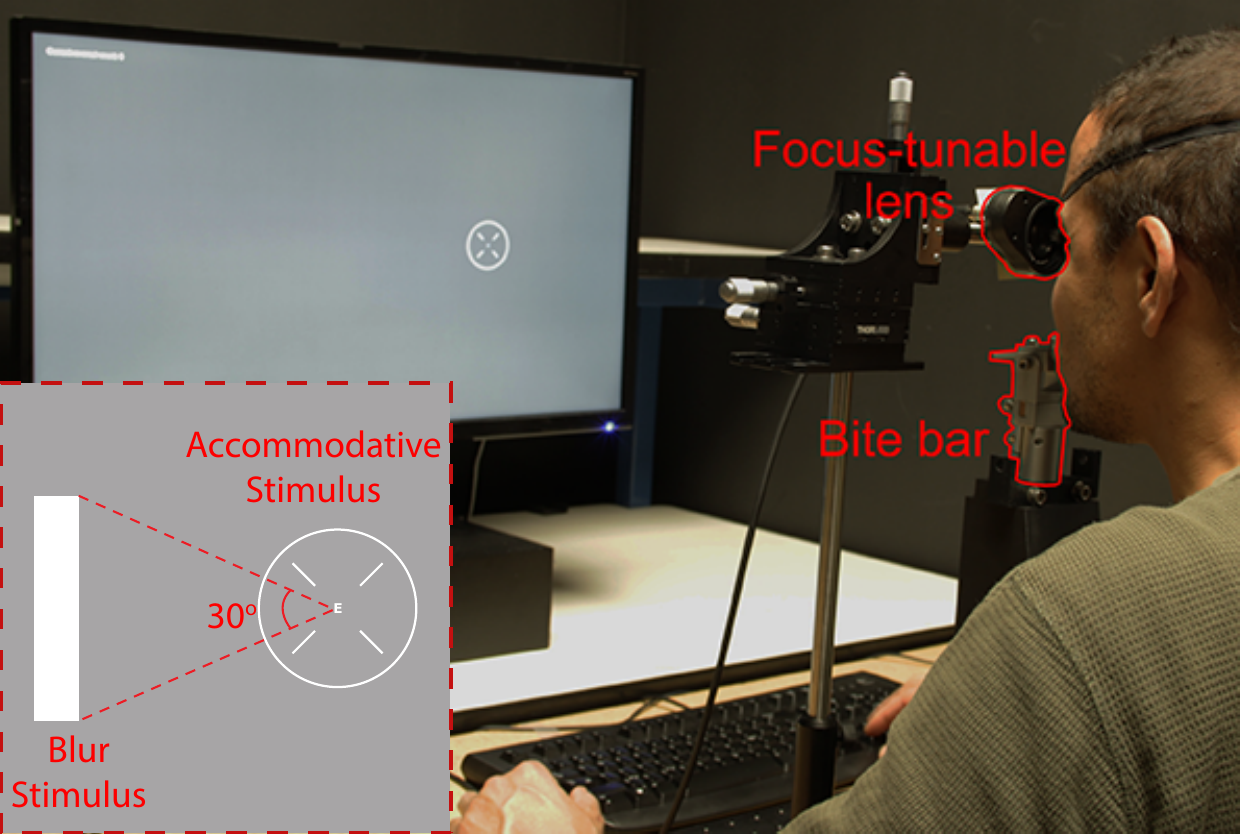}
    \caption{Blur perception study setup and a sampled stimulus (see \textcolor{urlblue}{Visualization 1}).}
    \label{fig:blur_experiment_design}
\end{figure}

Wang and Ciuffreda \shortcite{Wang:2006:EZF} measured the thresholds of blur detection (noticing blur) and discrimination (differentiating between blur sizes) at various visual eccentricities from fovea to $8\deg$.
The thresholds increased monotonically along with eccentricity for both detection ($0.53\diopter \rightarrow 1.25\diopter$) and discrimination ($0.29\diopter \rightarrow 0.72\diopter$). \nothing{$\diopter$ is the diopter unit, which differs from \emacsquote{dimension} in \emacsquote{4D} and \emacsquote{2D} displays.}
Ronchi and Molesini \shortcite{Ronchi:1975:DOF} also measured blur detection threshold at farther eccentricities, where it increased from $2$ to $5\diopter$ at $7\deg$ of eccentricity to $7$ to $12\diopter$ at $60\deg$ of eccentricity.
The monotonic increases suggests that we may only weakly perceive blur and depth at far periphery.

We note the wide disagreement in measured thresholds at $7$ - $8\deg$ of eccentricity between the two studies and measure the blur detection and discrimination thresholds from fovea to periphery (at $0$, $5$, $10$, and $15 \deg$ of visual eccentricities).
\paragraph{Setup}
The setup is photographed in \Cref{fig:blur_experiment_design}.
Blur pedestals (the baseline for discrimination tasks) of $-2, -1, 0, 1,$ and $2\diopter$ were tested across eccentricities.
The stimulus was presented on an LCD display (Acer XB270HU, $2560 \times 1440$ resolution, $144Hz$ refresh rate) located at $80$ cm away from the viewer, where the central pixel subtended 1 arcmin from the eye.
A focus-tunable lens (Optotune, EL-16-40-TC, response time $30ms$), placed before one eye of the subject (the other covered by mask), controlled the focal distance of the stimulus.
The field of view (FoV) provided by the lens subtended $25\deg$ in diameter.
A bite bar was used to precisely position the viewer at the desired location.
\paragraph{Calibration}
Every subject went through a calibration procedure before starting the experiment.
We first measured the far point of accommodation by using a tumbling E test~\cite{Taylor:1978:TE} and a staircase procedure~\cite{Levitt:1971:SC}.
Second, we quantified scaling and translation caused by the change in lens focal power using alignment tasks. These two calibration steps let us present stimuli accurately in terms of focal power, visual size, and location in the visual field.
\paragraph{Stimuli}
As shown in the bottom left inset of \Cref{fig:blur_experiment_design}, the visual stimulus for blur detection/discrimination was a bright rectangle ($100 {cd}/{m^2}$) drawn on a dark background ($20 {cd}/{m^2}$).
The size of the foveal rectangle was $0.16$ (W) $\times$ $0.8$ (H) $\deg$.
The peripheral rectangles scaled linearly with visual eccentricity: $0.04E$ (W) $\times$ $0.2E$ (H) $\deg$, where $E$ is the visual eccentricity in deg.
The focus-tunable lens operated with the display to introduce defocus blur to the rectangles.
The presentation time for the rectangle was $0.3\sec$, short enough to prevent human refocus \new{($0.3-0.4\sec$ according to \cite{Campbell:1960:DAR})}.

The rectangle appeared twice in a random sequential order with different amounts of blur; one with pedestal blur serving as a reference (or no blur for detection) and the other with blur greater or less than pedestal serving as a test signal. We define the difference between pedestal and test blurs as the differential blur. The magnitude of differential blur was adapted based on the 1-up 2-down staircase method with the minimum step size of $0.1\diopter$. The two stimuli were separated by 0.5 sec in time.
A fixation target (the letter \emacsquote{E} in the circle as marked in \Cref{fig:blur_experiment_design}) was presented before, between, and after the two blur discrimination stimuli to keep subjects' accommodative state constant. inserted for $0.5\sec$ between the two intervals to discourage subjects from accommodating to the stimulus.
The task was a 2-alternative-forced-choice: subjects had to choose the one that appeared blurrier even if they were not sure.

More than 100 trials were executed per each combination of blur pedestal and visual eccentricity.
The total duration, including calibration, training, and breaks, was about 6 hours. A capture of the study can be seen from \textcolor{urlblue}{Visualization 1}.
\paragraph{Subjects}
Four subjects, aged 31 to 48, participated.
All subjects had normal or corrected-to-normal visual acuity.
One was an author.
The other three  were unaware of the experimental hypothesis. All subjects provided written consent, and the experiment was conducted in accordance with the Declaration of Helsinki.
\begin{figure}[!th]
  \centering
  \includegraphics[width=\linewidth]{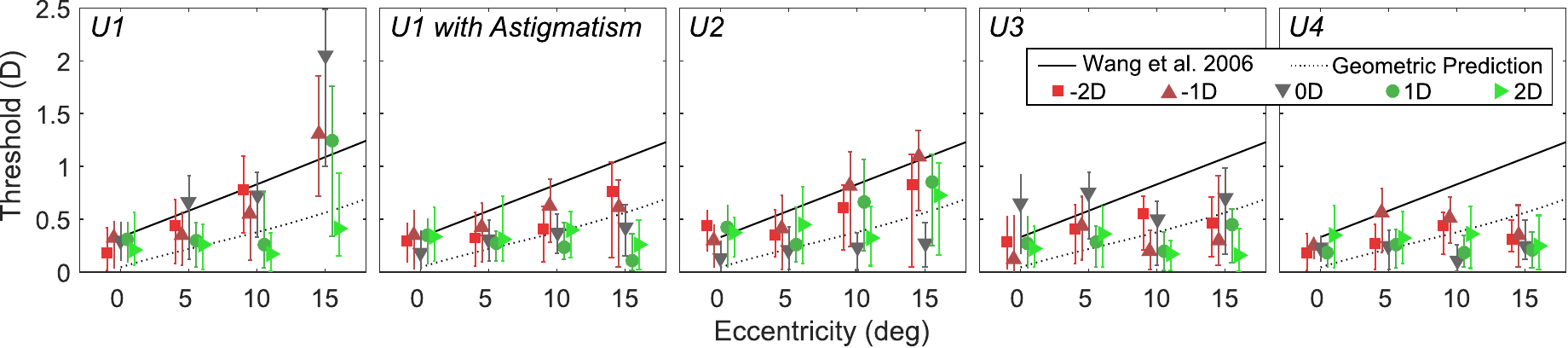}
  \caption{Blur study results. The Figs. plot the blur detection/discrimination thresholds as a function of eccentricity and pedestal/baseline blur ($-2, -1, 0, 1, 2\diopter$) for four subjects.
X-axis represents retinal eccentricity in degree.
Y-axis represents measured thresholds in diopter.
Each vertical bar indicates the $75\%$ performance level centered at a $95\%$ confidence interval.
See \textcolor{urlblue}{Data File 1} for underlying values.}
    \label{fig:blur_experiment_results}
\end{figure}
\paragraph{Results}
\label{sec:pilot_study:blur:results}
As in \Cref{fig:blur_experiment_results}, the Y-axis represents blur discrimination thresholds.
Each symbol indicates the $75\%$ performance level centered at a $95\%$ confidence interval. We define $75\%$ performance level as the threshold, which is halfway between random guess (there were two choices) and being perfect.
We estimated the threshold by fitting a cumulative Gaussian function to the performance curve drawn as a function of differential blur sizes \cite{Schutt:2016:PAB}. 
Specifically, we estimate participants' performance levels as a function of the added blur magnitude.
\nothing{Note that the performance level is not equal to whether users can tell the difference: it is $50\%$ (chance level) when they cannot, and $100\%$ when they can. 
The threshold is defined as the blur magnitude where $75\%$ performance occurs. We used a statistical method, psychometric function fitting \cite{Schutt:2016:PAB}, to estimate it.}

The experimental results are plotted in \Cref{fig:blur_experiment_results}, where pedestal blur is color coded. Note that $0\diopter$ pedestal means detection.
The dashed black line shows geometrically estimated detection thresholds by considering a cylindrical blur kernel, a $5mm$ pupil and anatomical ganglion cell densities from \cite{Watson:2014:FHR}.
The solid black line shows the prediction curve suggested by \cite{Wang:2006:EZF}.
Both lines are for theoretical comparison with our data.
The results show three observations.
First, the thresholds increased along with eccentricity for some subjects but not all; the values from two subjects (U3 and U4) remained nearly constant and below the theoretical predictions at farther eccentricities.
Second, correction for astigmatism of one subject (U1, astigmatism$=-1.25\diopter$ at $8\deg$) did not significantly improve, if not reduced, sensitivity to blur.
Third, varying pedestals sometimes yielded large threshold differences at the same eccentricity, although the hypothesis is a negative correlation \cite{Wang:2006:EZF}.
This may be attributable to individual differences such as peripheral refractive state \cite{Seidemann:2002:PRE}.
\section{Depth perception and visual eccentricity}
\label{sec:pilot_study:depth}
\label{sec:pilot_study:min_disparity}
We perceive depth using various visual cues where blur is \cite{held2012blur} one of them. Here we measured depth discrimination thresholds. \nothing{The stimulus used in this experiment contained  visual stimulus providing not only blur but also color \cite{Troscianko:1991:RCM} and spatial patterns \cite{Neri:2011:CFD}. Pictorial cues such as perspective \cite{van1979interrelation} was
To study the depth perception beyond only optical blur, we choose to experiment with a light field display generating natural defocus blur (see \Cref{fig:pilot_study:foveation}).}
We presented the stimulus on a 4D light field display, which can vary the intensity of light not only spatially but also angularly \cite{Ives:1902:ANS,lanman2013near}.
\nothing{We first conducted a pilot study showing the degraded depth perception as eccentricity grows, and then determined the detection threshold. To our best knowledge, no prior work studied the effect of eccentricity on depth perception using light field displays.}
\paragraph{Setup}
Figure \ref{fig:pilot_study:foveation} shows the study setup. We used a parallax-barrier-based \cite{Ives:1902:ANS} \new{($300 \mu m$ pitch size and $120 \mu m$ pinhole aperture)} light field display prototype built with off-the-shelf components (\cite{Sun:2017:PFL}). \new{The display panels was built by horizontally tiling three 5.98-inch 2K panels (Topfoison TF60006A) and a 3D printed housing. }It supports $3.2$ views$/\deg$ \new{angular and $579\times 333$ spatial resolutions}. \new{Comparing with the designs suggested by previous literature \cite{Takaki:2006:HDD,Huang:2015:LFS}, the display maintains high angular resolution to provide precise accommodation to validate our study result.}

During the study, subjects remained seated at $30cm$ ($15\deg$ FoV) from the display with their non-dominant eye occluded. A chinrest was used to precisely control the viewing distance. In this configuration, each pixel in the light field display provides two views both horizontally and vertically for the smallest pupil ($2 mm$ in diameter).
\begin{figure}[thb]
  \centering
  \subfloat[defocus]{
    \includegraphics[width=0.48\linewidth]{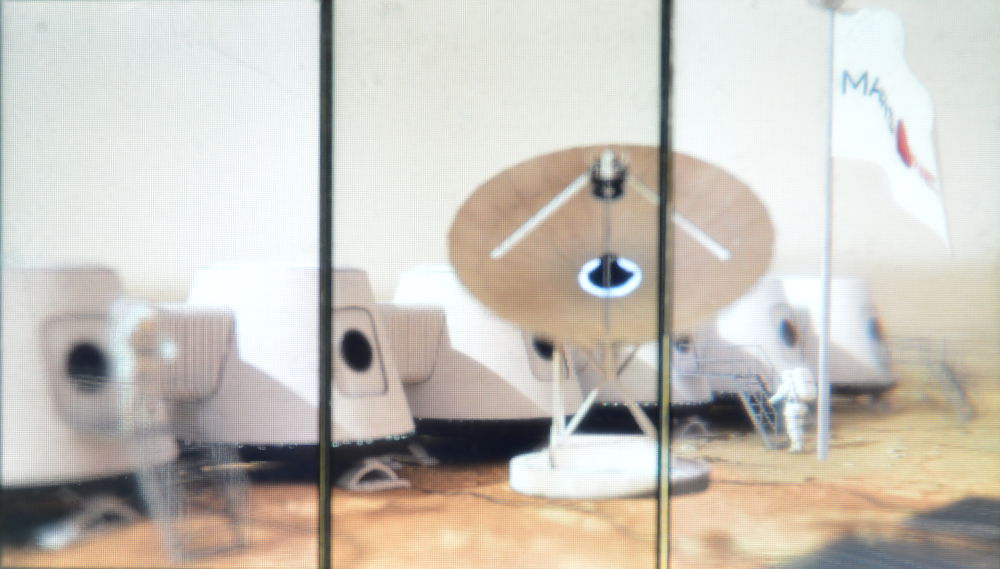}
    \includegraphics[width=0.48\linewidth]{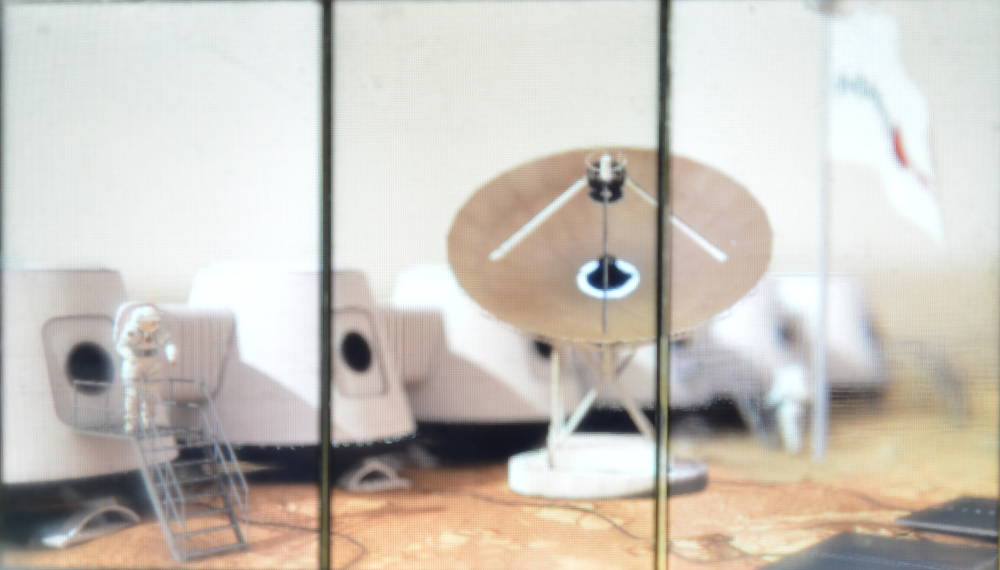}
    \label{fig:pilot_study:foveation:disparity}
  }%
  \hspace{1ex}

  \subfloat[setup and stimuli]{
    \label{fig:pilot_study:foveation:user}
    \includegraphics[width=0.6\linewidth]{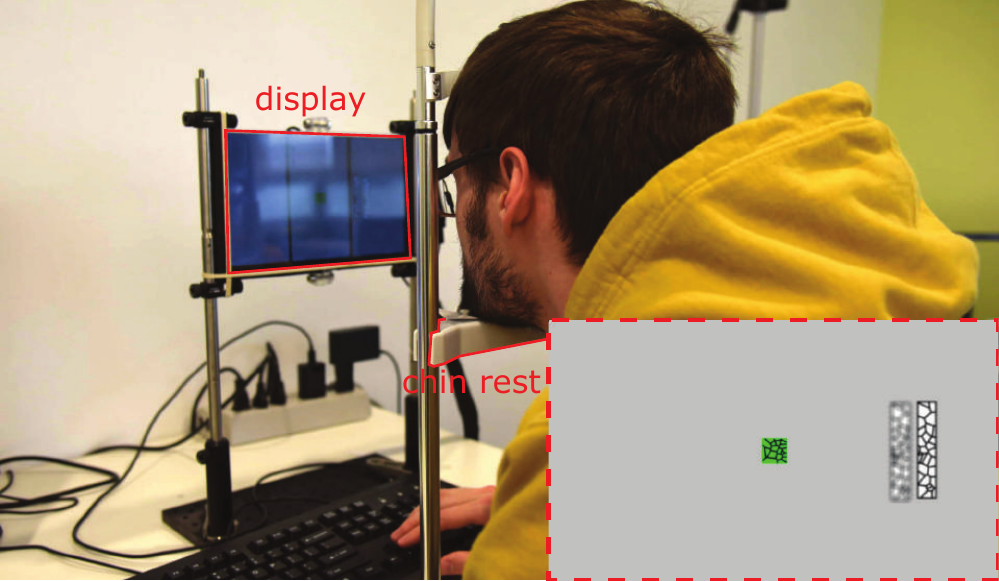}
  }%
  \caption{Depth perception study design. 
  \new{\protect\subref{fig:pilot_study:foveation:disparity} shows simulated retinal images via DLSR camera capturing the display \cite{Sun:2017:PFL}. The camera lens focus changed from far (left) to near (right) depth.}
  \protect\subref{fig:pilot_study:foveation:user} shows the study setup. The bottom inset shows a simulated retinal image from the stimuli. The green object is the fixation; the other two are the test targets (see \textcolor{urlblue}{Visualization 2}).}
  \label{fig:pilot_study:foveation}
\end{figure}
\nothing{
\begin{figure}[thb]
  \centering
  \subfloat[defocus]{
    \includegraphics[width=0.42\linewidth]{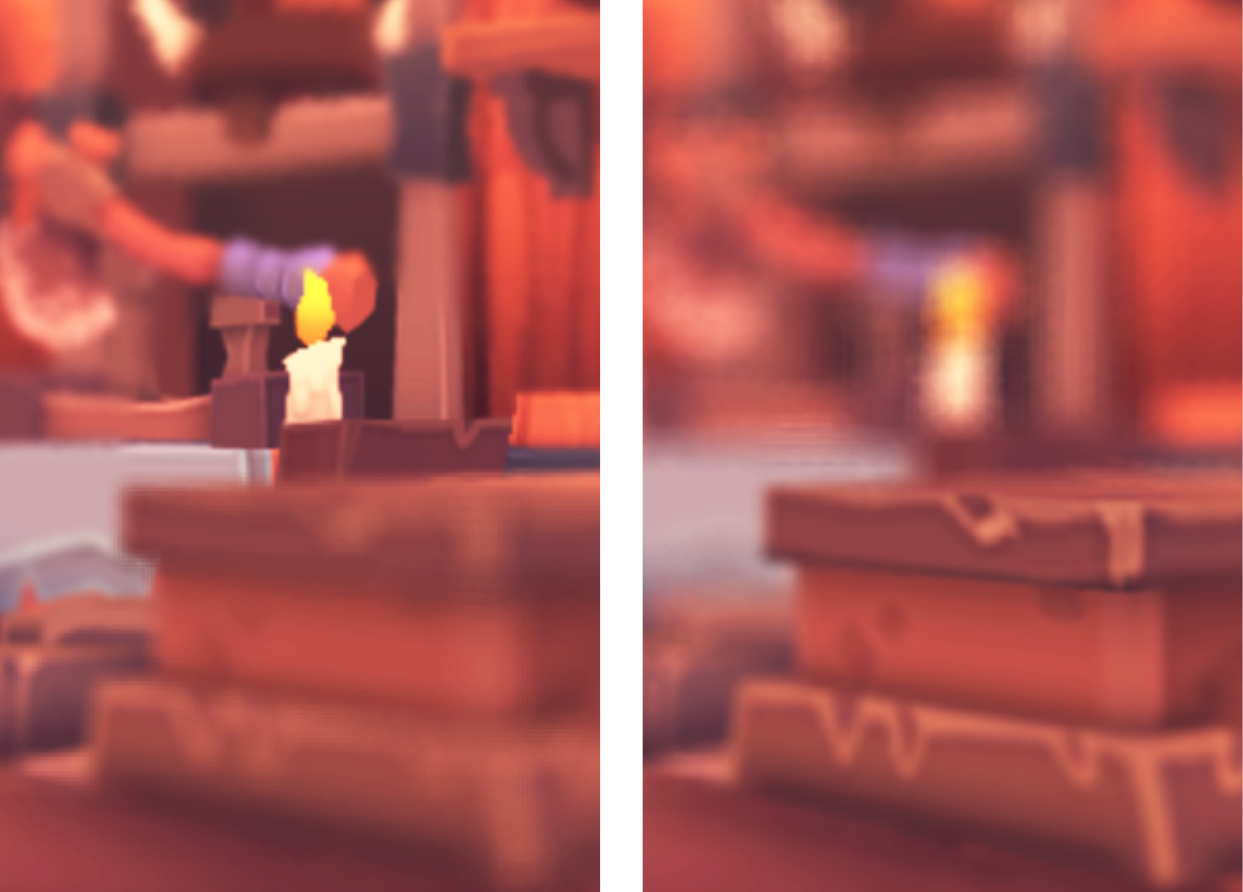}
    \label{fig:pilot_study:foveation:disparity}
  }%
  \hspace{1ex}
  \subfloat[setup and stimuli]{
    \label{fig:pilot_study:foveation:user}
    \includegraphics[width=0.46\linewidth]{user_stimuli.pdf}
  }%
  \caption{Depth perception study design. 
  \protect\subref{fig:pilot_study:foveation:disparity} shows simulated retinal images while the users are watching our display and changing focus from far (left) to near (right) depth \cite{Sun:2017:PFL}.
  \protect\subref{fig:pilot_study:foveation:user} shows the study setup. The bottom inset shows a simulated retinal image from the stimuli. The green object is the fixation; the other two are the test targets (see \textcolor{urlblue}{Visualization 2}).}
  \label{fig:pilot_study:foveation}
\end{figure}
}
\paragraph{Stimuli}
The bottom inset in \Cref{fig:pilot_study:foveation}(b) shows a sampled stimuli. The subjects were instructed to fixate on the fixation target.
The fixation target at \delete{$3.8\diopter$}\new{$3.2\diopter$} was a small green square, which remained at the the screen center to fix subjects' gaze and focal distance. 
On the side were two depth discrimination targets with broadband binary Voronoi diagrams textures on them. The test targets were two vertically elongated rectangles rendered side-by-side with a small $2.5mm$ gap to avoid cues from occlusion while keeping their eccentricities close to the tested condition.
To avoid size cue, we dynamically rescaled the sizes of the test targets based on their depths so that they always appear the same size in visual angle.
Subjects were instructed to keep watching the fixation target during the entire study. 
\nothing{In the first experiment, 
one test target was rendered at $3.8\diopter$ depth (same as the fixation), while the other at $2.8\diopter$. 
They were positioned together at either fovea or $13\deg$ periphery.
We ran 20 trials for each of these 2 conditions.
In each trial, subjects selected which target appeared closer.
The horizontal orders of the two targets were randomized across trials.}

\nothing{In the second experiment,}
The two targets appeared at one of $8$ eccentricities from the fovea to $15\deg$.
We used the method of adjustment to measure the depth detection thresholds. At the beginning of each trial, the two test targets were positioned at the same depth \new{as the fixation} ($3.2\diopter$). Then, subjects pressed up/down arrow buttons to increase/decrease the two targets' depth separations until reaching the thresholds where they perceive the relative depths. A warning appeared when the depth disparity reached $0$ or the hardware limit. We ran $4$ trials for each eccentricity.
All conditions were randomized. A capture of the study can be seen from \textcolor{urlblue}{Visualization 2}.
\paragraph{Subjects}
Four subjects, aged 23 to 46, participated. One was an author. The other three were unaware of the experimental hypothesis.
\paragraph{Results}
\nothing{The ratio of correct answers (fovea/periphery) in the first experiments were: $95\%/40\%$, $90\%/60\%$, $90\%/55\%$ and $95\%/55\%$ respectively for each user. The significantly higher accuracy in the fovea ($\geq90\%$) than periphery (near $50\%$, i.e., random guess) indicates a shared trend of foveated depth discrimination.}

\begin{figure}[thb]
    \centering
    \includegraphics[width=\linewidth]{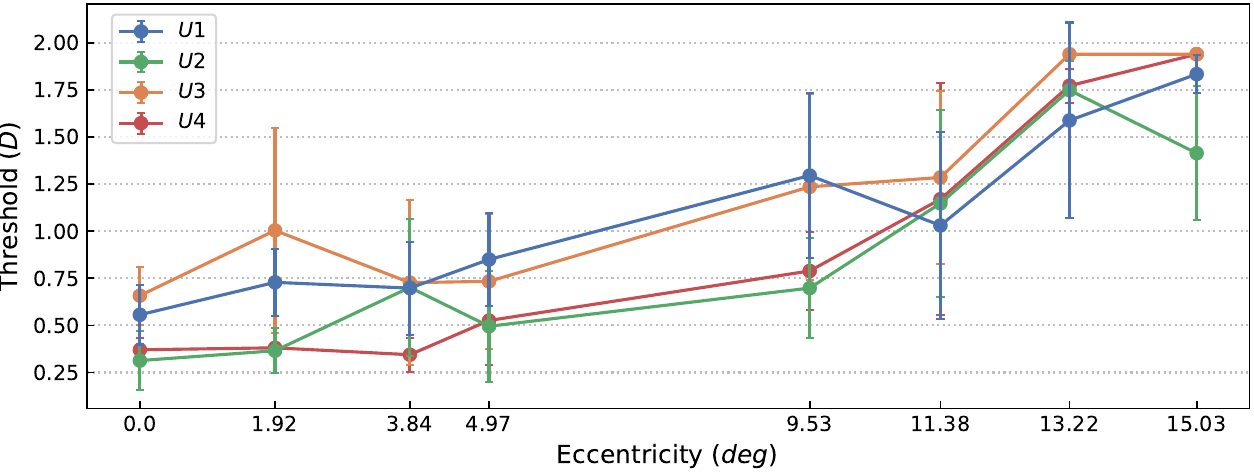}
    \caption{The result of depth detection thresholds (Y) against eccentricity (X). See \textcolor{urlblue}{Data File 2} for underlying values.}
    \label{fig:pilot_study:eccentricity:result}
\end{figure}

Figure \ref{fig:pilot_study:eccentricity:result} plots the mean thresholds with standard deviation\nothing{ in the second experiment}. 
All subjects showed consistent trend of increasing depth detection thresholds as eccentricity grows. We observed less degree of individual differences than we did in the first experiment on blur discrimination.
\section{Discussion}
\label{sec:pilot_study:discuss}
We found that blur discrimination threshold is significantly lower than previously reported. There two notable differences between our methods and those used previously. First, we used two alternative forced choice task, which might have helped subjects with performing at their best, where previous studies used method of adjustment. Second, our subjects used their habitual optical correction where previous studies had paralyzed subjects' accommodation with treatment \cite{Wang:2006:EZF}. We may discern the difference in blur better when harnessed with natural accommodative mechanisms. 

Nonetheless, the blur discrimination threshold reported in this study was surprisingly small. The geometrical calculation suggests that we can even discern the differences smaller than the spacing between retinal ganglion cells. Further investigation to clarify the mechanism for blur detection and discrimination will be beneficial.

Computational displays with depth cues and fast response time can benefit a variety of applications, such as in VR/AR and automotive assistance. 
To save computation like in resolutional foveation, we could leverage the consistently reduced monocular depth perception at larger eccentricity (\Cref{fig:pilot_study:eccentricity:result}\nothing{ show the consistency of decreased depth perception along eccentricities $\geq6\deg$\nothing{, although the exact values also vary among individuals}}).
{However,} we could yet do so for the commonly considered root cause, defocus blur.
The blur perception thresholds from optical stimulus (\Cref{fig:blur_experiment_results}) show that some individuals retain high blur sensitivity as far as $15\deg$, with the minimum threshold down to $0.2\diopter$. 
The different trends discovered from our two studies suggest that future research should analyze the whole retina-lens-display system: not only the optical/anatomical low-level vision but also the visual cortex processing.

\section*{Funding}
\new{This work has been partially supported by National Science Foundation grants IIP1069147, CNS1302246, NRT1633299, CNS1650499, and Hong Kong RGC general research fund 17202415.}


\section*{Disclosures}
The authors declare no conflicts of interest.



\bibliography{oe-reference}

\end{document}